# Elemantra: An End-to-End Automated Framework Empowered with AI and IoT for Tackling Human-Elephant Conflict in Elephant-Range Countries

Nuwan Sriyantha Bandara[1,2*], and Dilshan Pramudith Bandara[3*]

[1]Department of Electronic and Telecommunication Engineering, University of Moratuwa, Moratuwa 10400, Sri Lanka
[2]School of Computing and Information Systems, Singapore Management University, 80 Stamford Rd, Singapore
[3]Department of Electrical Engineering, University of Moratuwa, Moratuwa 10400, Sri Lanka
*Member, IEEE

*Abstract*—The cohabitation of elephants and humans has evolved into a human-elephant conflict (HEC) due to the increasing loss of historical elephant habitations. Since HEC is a substantial threat to both species, advanced sensing methods are utilized to develop HEC prevention frameworks which still lack unification. Here, we propose an end-to-end automated framework for HEC prevention consisting of three main modules: a distributed deep learning-assisted elephant detection module using infrared and seismic sensing, an on-site repelling system with time-varying acoustic and light deterrents and a mesh network for device communication. The framework is equipped with novel decision-making pipelines and algorithms such that it has the potential to operate with no human intervention. The preliminary results from each module confirm that the proposed framework is effective in performing their individual tasks towards collaboratively achieving the prevention of HEC. The codes are publicly available at https://github.com/nuwansribandara/elemantra.

*Index Terms*—Human-Elephant Conflict, Deep Learning, Internet of Things, End-to-End Framework, Resource-Constrained Environments

## I. INTRODUCTION

The cohabitation of elephants and humans in Asia and Africa has evolved over the past thousands of years due to the continuous competition for resources between the two species, especially wherever they share a landscape [1]. However, the gradual expansion of human settlements and agricultural fields has caused a significant shrinkage in historical elephant habitations, lesser landscape connectivity, reduced forage [2] and consequently a widespread deterioration in elephant population [3]. The increasing loss of both habitations and forage progressively forces elephants into closer contact with humans, leading to more frequent and destructive conflicts over scarce resources [4] which eventually results in severe consequences from crop-raiding to loss of lives of both species [3]. Therefore, human-elephant conflict (HEC) is undoubtedly a substantial threat to species' survival as well as to the social and ecological sustainability in elephant-range countries for which proper conflict management strategies are needed to ensure the coexistence between elephants and people [5]. In this context, a wide variety of HEC prevention and mitigation approaches have been developed and adopted recently, based on the deepened understanding of the nature and spatio-temporal patterns of HEC [2], such as exclusion and on-site deterrents as prevention techniques and elephant translocation and monetary compensation for losses as mitigation techniques [2].

Traditionally, HEC prevention is addressed through exclusionary methods such as protected ecological areas, electric fences and trenches or on-site deterrent methods such as acoustic, light-based and agriculture-based deterrents [2]. However, the recent advancements in sensing, computing and communications drive the development of early detection and warning systems as a promising HEC prevention technique [2], [6] in which potentially problematic elephants or herds are detected through the placement of sensors and repelled from conflict-prone locations [2]. Over the past few years, many works have been proposed to automatically detect elephants using a diverse array of sensors such as infrasonic sensors [7], passive infrared sensors [8], satellite and aerial imagery [9] or tracking [10] and seismic sensors [11] whereas repelling is mainly based on automated acoustic playbacks [12] or light-based deterrents [13]. However, most of these approaches focus on either detection or repelling alone or are still in need of considerable human intervention to ensure reliable detection and effective repelling due to their crucial design limitation of not being an end-to-end and fully automated framework, towards achieving the utmost goal of HEC prevention.

In this paper, we propose an end-to-end fully automated framework for HEC prevention in elephant-range countries with an enhanced detection and repelling mechanism. The proposed framework consists of three main modules: 1) a distributed, deep learning (DL)-assisted detection module that consists of hybrid sensing (i.e. seismic and infrared) and a collaborative decision framework, 2) an on-site repelling module that incorporates a combination of time-varying acoustic and light-based deterrents, and 3) a mesh network to connect peripheral sensing nodes with the central node (CN) in the network for data and decision signal communication. The proposed framework is designed with custom decision-making pipelines such that it has the capability to perform in an end-to-end fashion from detecting conflict-prone elephants to repelling such elephants with no human intervention. The framework is further integrated with an early warning system to alarm the people and officials to carry out proactive measures in the context of HEC.

## II. PROPOSED FRAMEWORK

The proposed framework, shown in Fig. 1, in its primary network configuration, consists of a CN which is responsible for infrared image-based elephant detection and peripheral nodes (PNs) which are tasked with seismic signal processing and the on-site repelling. Nodes are interconnected through a mesh network for device communication. The framework, given that it is positioned in a strategic manner, has the ability to maximally cover the conflict-prone landscape through the PNs. Further, it has the potential to semantically expand its coverage through the mesh network of nodes as per the specific requirements of the location.

When a conflict-prone elephant or a herd of elephants approaches the sensor-covered landscape, the PNs, which are positioned near the elephant congregating hot spots, sense the respective physical input that corresponds to the presence of elephants through the deployed seismic or infrared sensors. Then, the PN at the location or the CN processes the acquired sensing signals, via the decision framework assisted by DL or 1D signal processing, to detect the presence of elephants. If the presence is determined through the above framework, the on-site repelling system at the respective PN is activated to deter the elephants. Further, the wildlife officers and the near human habitats are notified for proactive intervention through mobile-based messages and alarming sirens respectively. Fig. 2 presents the complete operating algorithms.

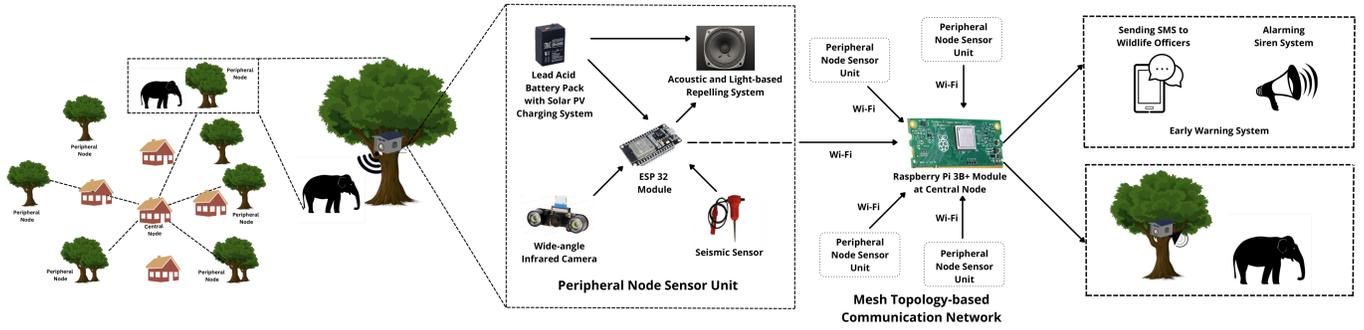

Fig. 1. Proposed framework overview: PNs, each with hybrid sensing, are positioned to sense elephants whereas CN and PNs collaboratively detect the elephants using acquired sensor signals. If detected, the on-site repelling mechanism at PNs and the early warning system is activated.

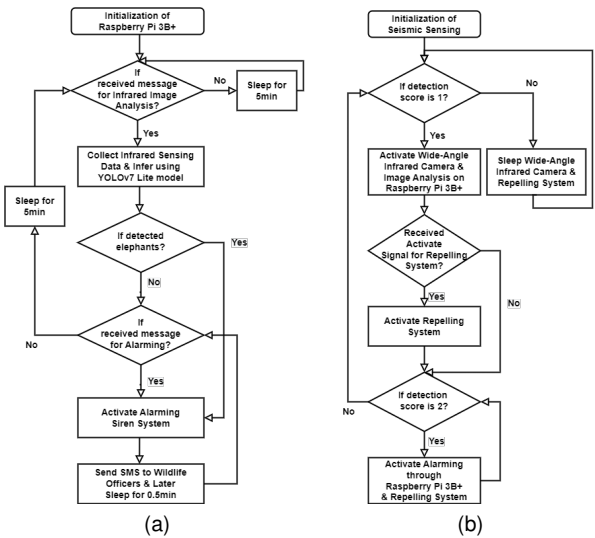

Fig. 2. Complete operating algorithms of the proposed framework at (a) central node (b) peripheral nodes

### A. Central Node

The CN of the proposed framework is responsible for: (1) processing the infrared images sent by the infrared cameras at PNs to detect elephants since the PNs do not constitute sufficient computational capacity to process such images in real-time and (2) communicating with PNs for data and decision transmission. The node is mainly comprised of a Raspberry Pi 3B+ (RPi-3) in our implementation for computing. CN is positioned in a human habitat such that it is connected to the grid for continuous power supply and all PNs within the network have the capability to communicate.

The acquired images by PN are sent to the CN where a DL-based model is deployed on the RPi-3. A tiny version of YOLOv7 [14] is developed to facilitate real-time processing on the resource-constrained edge device, RPi-3, without significantly compromising detection performance. When deriving the tiny version, the default hyperparameters are utilized as in [14] and the PyTorch-based pre-trained weights of the original model are deployed using a transfer learning strategy. The conversion process from the PyTorch model to the quantized Tensorflow (TF) lite model (in tflite format) includes two major intermediate steps: ONXX model and TF model. See section III-B for used datasets and results.

### B. Peripheral Nodes

The PNs are responsible for sensing the physical quantities corresponding to the elephants via the deployed seismic and infrared sensors and processing the acquired 1D seismic signals to determine the presence of elephants. The nodes are mainly comprised of an ESP32 unit for computing, a wide-angle infrared camera (i.e. infrared sensor), a geophone sensor (i.e. seismic sensor), a low-power speaker, a flashlight and the power unit.

In PNs, the seismic sensor monitors omni-directional ground vibrations caused by elephant movements or low-frequency vocalization rumbles [11], while the infrared sensor captures the thermal signatures of elephants. This combination of sensors ensures early detection with enhanced accuracy at both day and night times, enabling timely responses to potential conflicts. The seismic and infrared sensors are strategically placed to cover hot spots or areas, where elephants are known to congregate or walk through, enhancing the system's efficacy. Furthermore, by targeting these precise locations, the occurrence of false alarms in areas with lesser elephant activity can also be minimized.

As in Fig. 2, the infrared sensor is only activated when the detection score ($ds$) from seismic signal processing reaches a defined threshold in order to reduce the power consumption at PNs. The algorithm 1 is proposed to calculate $ds$ and it is developed, to run in the edge ESP32 with lower computational cost, based on the intuition that the seismic signals with 3-5 s duration and characteristic frequency modulation (i.e. initially increasing and then decreasing peak frequency between 20 Hz and 40 Hz) represents the elephant presence [11]. Further, it is to be noted that the acquired continuous seismic signals are preprocessed in the analog domain, sampled and divided into non-overlapping windows before feeding into the shown algorithm 1 which is for a $4s$ window of the processed signal.

The repelling system includes an acoustic mechanism using a low-power speaker for replicating bee sounds to deter elephants from approaching human territories. This approach leverages elephants' instinctive aversion to bees [15], creating a natural deterrent without causing harm or disruption to either elephants or humans. However, since it is evident that the elephants quickly learn to tolerate such sounds [16], we propose to modify the bee sounds either in the time or frequency domain, before releasing the sounds from the speaker, by randomly selecting either of the following three methods: (1) modification of the frame rate ($FR$) in the time domain where new $FR = \alpha \times FR_{old}$, (2) addition of an overlay of pink noise where the new sound is $y[n] = x[n] + N(\alpha, n)$, and (3) addition of an overlay of zero to indicate silent frames where the gap length of silence ($GL$) is $GL = \alpha \times 100ms$ and $\alpha$ is the uniform random factor. The repelling hardware also includes a unique flashlight. This light dims at a specific frequency when activated if an elephant presence is detected.

**Algorithm 1** Algorithm for seismic signal-based elephant detection
1: **Input:** Sampled windowed seismic signal **x**[n]
2: **Output:** The detection score $ds$
3: detected_indices ← [ ]; $ds$ ← 0; segment ← FFT{**x**[n]}
4: segment_freq ← FFT_FREQ{len(**x**[n]), $\frac{1}{sample\_rate}$}
5: $max\_freq$ ← **segment_freq**[INDEX$_{max}$[absolute(**segment**)]]
6: **if** $max\_freq$ is greater than 20 and less than 40 **then**
7:     **detected_indices**.append(1)
8: **else**
9:     **detected_indices** ← [ ]
10: **end if**
11: **if** len(**detected_indices**) is greater than 6 and less than 24 **then**
12:     $ds$ ← 1
13: **else if** len(**detected_indices**) is equal or greater than 24 **then**
14:     $ds$ ← 2
15: **end if**

The designated power source for each PN includes a cost-effective lead-acid battery pack in conjunction with a solar photovoltaic (PV) charging system. The battery pack stands as a steadfast energy storage solution, guaranteeing uninterrupted operation even during instances of diminished solar energy generation while the utilization of solar PV units for the PNs emerges as a fitting solution to meet the power requirements given the unavailability of grid power in the areas in proximity to forested regions. By harmonizing these elements, the power source of the system attains both robustness and environmental conscientiousness, crucial for sustainable operation.

### C. Mesh Communication Network

The PNs need to wirelessly communicate with the CN to send the infrared images and to receive the commands for repelling if an elephant presence is detected through infrared images. The intuition behind selecting a mesh topology is to enable the possibility of expanding the network through ESP32s as suitable for the landscape and to facilitate adding backup RPi-3s to the network if the connection between the main RPi-3 and ESP32s is interrupted. In our implementation, RPi-3 is the MQTT broker and ESP32s act as subscribers and publishers to the broker, based on WiFi for communications.

## III. EXPERIMENTAL RESULTS AND DISCUSSION

### A. Overall System

The proposed framework is compared with the existing state-of-the-art (SOTA) systems in Table 1 focusing more on the sensing features and unification. In the Table 1, if only a task, such as detection, is intelligently automated, it is ticked and so on.

### B. Elephant Detection

For experimental validation of the proposed DL model for elephant detection using infrared images, we developed a custom dataset, using the dataset presented in [18] and an in-house JavaScript script for web scraping, due to the absence of a diverse public dataset of infrared image-based animal detection. All the images in the accessed datasets were not considered for labelling because of the high annotation cost and the final dataset, which is used for training and inferring the DL model, contains 537 images in total in which approximately 75% of the total images has at least one elephant in it. The images from [18] are binary thresholded using Otsu's method, flood-filled and inverted before adding to the final dataset. All the collected images in the final dataset are annotated for elephants using the LabelImg tool [19].

In our implementation, the weights of the standard YOLOv7 model from [14] are transferred to initialize the training on our final dataset which is partitioned into 70%, 20% and 10% as training, validation and test datasets respectively. The trained PyTorch model is then converted to the tflite model to be saved and transferred to the RPi-3 device. The performance of the tflite model compared to the standard model is evaluated using average precision at the intersection-over-union (IoU) threshold of 50% (AP50) where $IoU = \frac{\text{area of intersection between the boxes}}{\text{area of union of the two boxes}}$ given the predicted bounding box and true bounding box.

TABLE 1. Comparison of proposed framework with SOTA systems

| Method | Sensing | Detection | Repelling | Networked | Warning |
|---|---|---|---|---|---|
| Thuppil [12] | infrared | ✗ | ✓ | ✗ | ✗ |
| Reinwald [11] | seismic & acoustic | ✓ | ✗ | ✗ | ✗ |
| Pemasinghe [17] | optical images | ✓ | ✓ | ✗ | ✗ |
| **Elemantra (Ours)** | seismic & infrared | ✓ | ✓ | ✓ | ✓ |

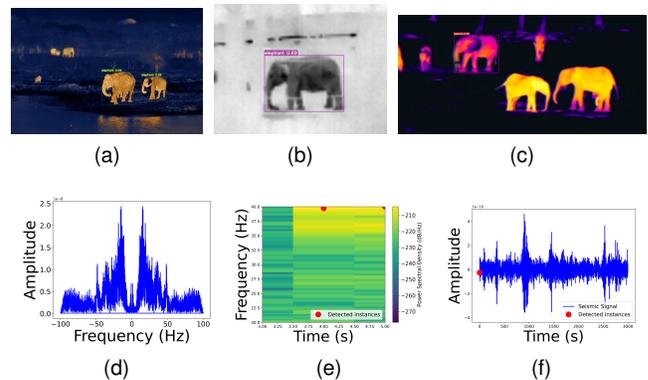

Fig. 3. First row: Elephant detection instances in test infrared set using tflite model (a) with high confidence, (b) one retrieved from [18], (c) one which failed to detect multiple elephants. Second row: Elephant detection instance using seismic recordings (d) frequency distribution, (e) detection in STFT plot, (f) detection using proposed method

In order to validate the proposed algorithm in section II-B for detecting elephants using seismic signals, we evaluate the approximation accuracy between the short-time Fourier transform (STFT) and the proposed method since STFT is capable of satisfactorily modelling the characteristic behaviour of seismic signals upon elephant movements or vocalizations as shown in [11]. Here, we utilize the acoustic and seismic dataset, which is developed for localization of elephant rumbles [11] and the dataset contains 44 seismic recordings in MSEED format which were collected using seismic field sensors buried within 70 cm deep in the soil.

In terms of the results from infrared images, the PyTorch and tflite YOLOv7 models achieve AP50s of 0.8952 and 0.6752 respectively. The superior performance from the standard model is expected since it has more parameters with higher floating point precision and model size than the tflite model which is specifically optimized and quantized to run on n RPi-3. However, the detection performance of tflite model is in the acceptable region which indicates a higher precision and recall in elephant detection (see Fig. 3). Furthermore, the performance of the proposed seismic signal-based algorithm is evaluated by measuring the recall of the proposed method compared to the STFT method using $recall = \frac{\text{number of corresponding instances detected by the proposed method}}{\text{number of instances detected by STFT}}$. As per the results from the used dataset, our method achieves a $recall$ of 0.82 which we believe is a significantly accurate approximation given the resource constraints in ESP32.

### C. Hardware Design

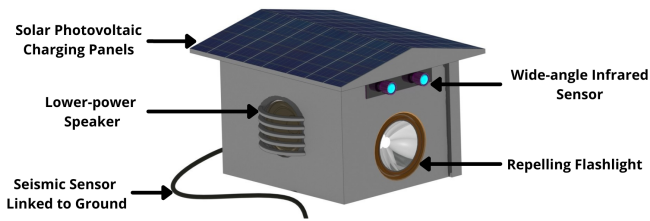

Fig. 4. Hardware design of the peripheral node

Our hardware design for the PN, depicted in Fig. 4, is designed using SOLIDWORKS software and combines essential components for the proposed PN. The sensor unit's seismic and infrared sensors pinpoint towards the identified hot spots whereas the integrated repelling hardware emits bee sounds and flashing lights triggered by sensor data. Solar PV roofs in PNs ensure reliable power in remote areas and therefore, the proposed design harmonizes technology and nature for sustained operations.

### D. Repelling System

To evaluate the proposed method for modifying the bee sounds to tackle the issue of elephants' ability to tolerate repetitive acoustic deterrents, we utilize the dataset presented in [20], developed for the automatic recognition of beehive sounds. The dataset consists of 78 recordings of varying lengths which approximately accumulate for a total duration of 12 hours and it is temporally segmented and labelled into two classes: Bee or noBee based on the perceived sound signal source being internal or external to the hive.

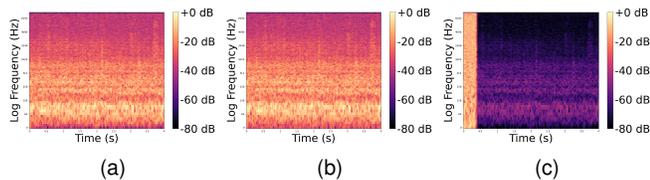

Fig. 5. STFT spectrogram plots of one (a) original bee sound segment of $4s$, the corresponding modified bee sounds through randomly (b) changing the *FR* and (c) adding an overlay of pink noise

The proposed methods for modification are tested using Python and then implemented in ESP32 connected to a low-power speaker for practical validation, which proved to be feasible. The performance of the proposed approach is further evaluated using the STFT plots as shown in Fig. 5 and the maximum cross-correlation scores between STFT plots which convey that the utilized approach is capable of modifying the bee sounds while having similar acoustic characteristics to original bee sounds in overall.

## IV. CONCLUSION

The proposed framework achieves enhanced reliability and accuracy in timely proactive interventions against elephant intrusions through novel algorithms for collaboratively detecting the elephants both day and night, while the integration of the proposed repelling mechanism offers a humane deterrent while being promising towards building sustainable automated acoustic deterrents against conflict-prone elephants. Therefore, through our framework, we innovatively address immediate conflict concerns towards fostering coexistence between humans and elephants.